\documentclass{aa}

\usepackage{graphicx}
\usepackage[varg]{txfonts}
\usepackage{natbib}
\usepackage[pdftex]{hyperref}

\graphicspath{{./figures/}}

\newcommand\figI{
  \begin{figure}
    \centering
    \includegraphics[width=8.5cm]{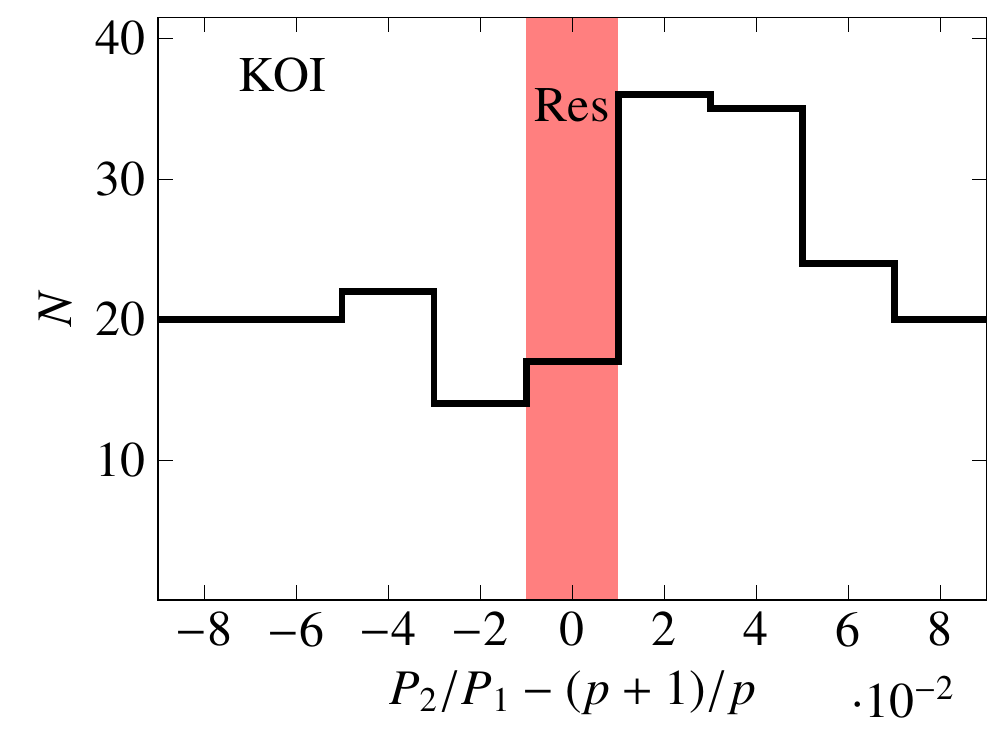}
    \caption{Distribution of period ratio between pairs of planets
      close to the 2:1 and 3:2 mean-motion resonances.
      The distributions around both resonances are accumulated in order to obtain a more
      important set of systems.
      These statistics are obtained from the Q1-Q16 KOI catalog \citep{batalha_planetary_2013}.
      The origin of the $x$-axis is the exact commensurability of the periods (resonant systems)
      and is highlighted with a red strip.
      Negative values correspond to internal circulation ($P_2/P_1<(p+1)/p$) and
      positive values correspond to external circulation ($P_2/P_1>(p+1)/p$).
      We observe an important excess of systems in external circulation, with
      $P_2/P_1 - (p+1)/p \approx 2\times 10^{-2}$.
    }
    \label{fig:I}
  \end{figure}
}

\newcommand\figII{
  \begin{figure}
    \centering
    \includegraphics[width=8.5cm,trim = 0cm 1.3cm 0cm 0cm, clip]{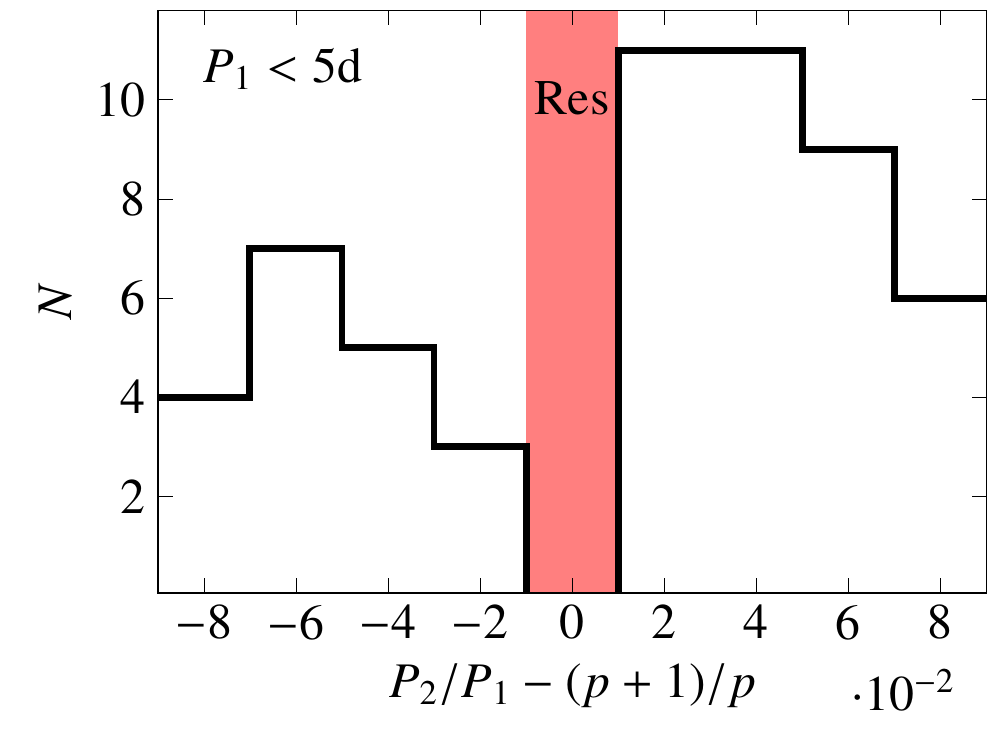}\\
    \includegraphics[width=8.5cm,trim = 0cm 1.35cm 0cm 1mm, clip]{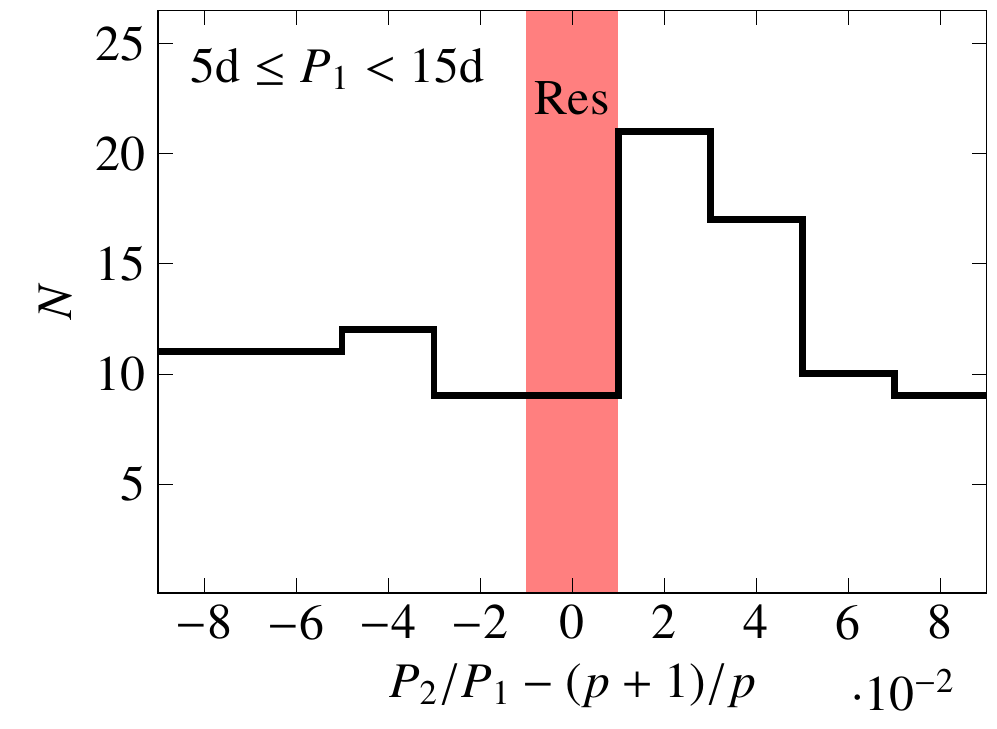}\\
    \includegraphics[width=8.5cm]{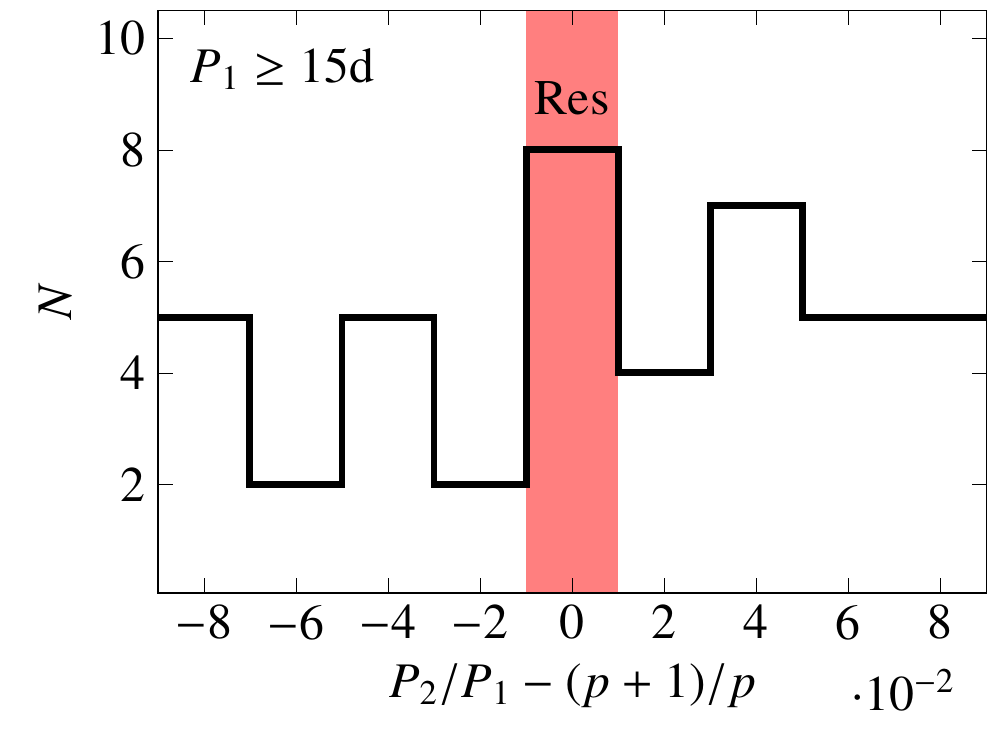}
    \caption{Same as Fig.~\ref{fig:I} but the statistics are computed using
      different subsets of KOI pairs depending on the period of the inner planet ($P_1$).
      We divide the data set in three groups: $P_1 < 5$d (\textit{top}),
      $5\mathrm{d} \leq P_1<15$d (\textit{middle}) and $P_1 \geq 15$d (\textit{bottom}).
      For the innermost systems (\textit{top}), we observe an important excess of planets in
      external circulation ($P_2/P_1 > (p+1)/p$ for the resonance $p+1$:$p$)
      and no resonant systems ($P_2/P_1 \approx (p+1)/p$).
      For the intermediate group (\textit{middle}), the excess of planets in external circulation
      is still visible but less important and a significant number of resonant planets are observed.
      Finally for the farthest systems (\textit{bottom}), the number of resonant planets is slightly higher
      than the number of planets in external circulation.
    }
    \label{fig:II}
  \end{figure}
}

\newcommand\figIII{
  \begin{figure}
    \centering
    \includegraphics[width=8.5cm]{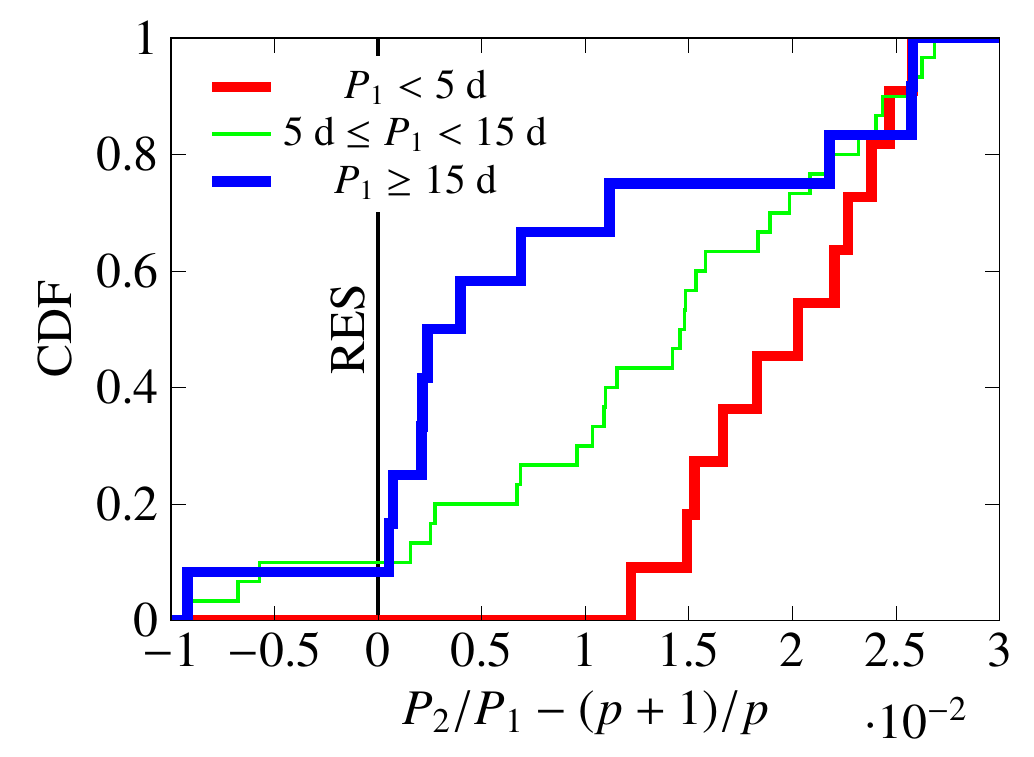}
    \caption{Cumulative distributions of planet pairs in the vicinity of the 2:1 and 3:2 mean-motion resonances
      (the statistics of both resonances are accumulated) for the three groups defined in Fig.~\ref{fig:II}
      (see also Sect.~\ref{sec:distance})
      The conclusions are the same as in Fig.~\ref{fig:II}: for farthest systems (blue)
      there is a pile-up of planets in the resonance, while for close-in systems (red)
      the pile-up is shifted toward
      higher values of the period ratio and we observe a lack of resonant systems.
      The distribution of intermediate systems (green) is, not surprisingly, intermediate.
      Using K-S tests to compare these distributions, we obtain a p-value of $0.08\%$ for both extreme
      distributions (red and blue).
      The p-value for the blue and green distributions is $10\%$,
      and for the green and red ones $3.5\%$.
    }
    \label{fig:III}
  \end{figure}
}

\begin{document}

\title{Tidal dissipation and the formation of Kepler near-resonant planets}
\titlerunning{Kepler near-resonant planets}
\author{J.-B. Delisle \and J. Laskar}

\institute{ASD, IMCCE-CNRS UMR8028, Observatoire de Paris, UPMC,
  77 Av. Denfert-Rochereau, 75014~Paris, France\\
  \email{delisle@imcce.fr}
}

\date{\today}

\abstract{
  Multi-planetary systems detected by the \textit{Kepler} mission present
  an excess of planets close to first-order mean-motion resonances
  (2:1 and 3:2) but with a period ratio slightly higher than the resonant value.
  Several mechanisms have been proposed to explain this
  observation. Here we provide some clues that these near-resonant systems
  were initially in resonance and reached their
  current configuration through tidal dissipation.
  The argument that has been opposed to this scenario is that it only applies to the
  close-in systems and not to the farthest ones for which the tidal effect is too weak.
  Using the catalog of KOI of the \textit{Kepler} mission,
  we show that the distributions of period ratio among the most close-in planetary systems
  and the farthest ones differ significantly.
  This distance dependent repartition is a strong argument in favor of the tidal dissipation scenario.
}


\maketitle

\section{Introduction}

The \textit{Kepler} mission has opened the opportunity
to perform statistical studies on a considerable number of planets.
More specifically, the large number of planets detected in multi-planetary systems allows to test
the formation and evolution scenarios of planetary systems.
One of the most surprising results obtained by the \textit{Kepler} mission
was the fact that only a small fraction of planet pairs are
locked in first-order mean-motion resonances (2:1, 3:2) whereas a significant excess of pairs is found with
a period ratio close to but higher than the resonant value
\citep{lissauer_architecture_2011,fabrycky_architecture_2012}.
We reproduce in Fig.~\ref{fig:I} the distribution of period ratio of planet pairs close to these first-order resonances
using the Q1-Q16 KOI catalog \citep{batalha_planetary_2013}.
This data set contains the \textit{Kepler} confirmed planets as well as
unconfirmed planet candidates.
Candidates that are known to be false positives are removed from the sample.
We observe, as described in the literature, an excess of planet pairs with
a period ratio higher than the resonant value (see Fig.~\ref{fig:I}).

\figI

Different explanations for this observation have been proposed involving tidal dissipation raised
by the star on the planets
\citep{lithwick_resonant_2012,delisle_dissipation_2012,delisle_resonance_2014,batygin_dissipative_2013},
dissipative effects between the planets and the proto-planetary disk
\citep{rein_period_2012,baruteau_disk_2013},
between planets and planetesimals \citep{chatterjee_planetesimal_2014},
or in-situ formation \citep{petrovich_planets_2013,xie_asymmetric_2014}.
In this article we provide some statistical clues in favor of the scenario
involving tidal dissipation in planets initially locked in resonance.

The phenomenon of resonant departure induced by tidal dissipation was described
by \citet{papaloizou_dynamics_2010} and \citet{papaloizou_tidal_2011} and has been
analyzed with a particular focus on \textit{Kepler} statistics by different authors
\citep{lithwick_resonant_2012,delisle_dissipation_2012,batygin_dissipative_2013}.
These studies showed that for close-in planetary systems an excess of planets similar
to the observed one is naturally produced
by tidal dissipation raised in the planets by the stars.

Recently, \citet{lee_kepler_2013} showed that this scenario
is too slow to explain the typical distance
of planet pairs to the nominal resonance ($P_2/P_1 - (p+1)/p \approx 2\times10^{-2}$)
on a reasonable timescale.
In \citet{delisle_resonance_2014}, we showed that tidal dissipation raised by the star in the innermost planet
induces an increase of the amplitude of libration in the resonance.
If the initial amplitude of libration (at the time of disappearing of the proto-planetary disk) is significant,
the system is able to cross the separatrix and leave the resonance while the eccentricities of the planets are
still important ($e_1\gtrsim 0.15$).
The subsequent evolution of the period ratio of the planets is in this case
3-5 orders of magnitude higher than in the scenario of departure at low eccentricities
considered by \citet{lee_kepler_2013}
\footnote{\citet{lee_kepler_2013} considered the same scenario of resonance departure at low eccentricities
  as in previous studies
  \citep{papaloizou_dynamics_2010,papaloizou_tidal_2011,lithwick_resonant_2012,delisle_dissipation_2012,batygin_dissipative_2013}.
}
because the tidal effect gets more efficient with increasing eccentricities
\citep[see][section 5]{delisle_resonance_2014}.
Therefore, many systems that were discarded by \citet{lee_kepler_2013} could actually have evolved
from the resonance to their current configuration by tidal dissipation, following this new scenario.
Supposing an important initial amplitude of libration in the resonance is not absurd.
\citet{goldreich_overstable_2014} showed that during the phase of migration in the proto-planetary
disk, many planet pairs that are locked in resonance have their amplitude of libration increased
by the dissipation induced by the disk.

However, this new scenario still involves the tidal effect and should thus be very efficient
for close-in systems but not for the farthest ones.
This is the main argument opposed to the tidal dissipation scenario
\citep[e.g.][]{rein_period_2012,baruteau_disk_2013}.
On the contrary, the other proposed mechanisms do
not have an important dependency on the distance to the star.
In the following we reanalyze the \textit{Kepler} statistics with a focus on the distance of the planets
to the star.

\section{Dependency on the distance to the star}
\label{sec:distance}

Different authors already analyzed the impact of the distance to the star on the
distribution of systems close to first-order mean-motion resonances.
\citet{rein_period_2012} divided the sample of \textit{Kepler}
planet pairs in two groups depending on the period of the innermost planet.
The author used a threshold at 5 days and
found a similar distribution for systems with $P_1<5$d and for those with $P_1\geq 5$d.
Using a threshold at 10 days, \citet{baruteau_disk_2013} reached the same conclusion.
Both studies discarded the scenario of a tidally induced distribution of period ratio
since according to this scenario the excess should only be observed for the innermost systems.

\figII

In Fig.~\ref{fig:II} we show the results of a similar study on more recent data
\citep[Q1-Q16 KOI catalog,][]{batalha_planetary_2013}.
Our findings are in disagreement with previous studies.
We divide the data set in three groups:
\begin{enumerate}
  \item close-in systems with $P_1<5$d,
  \item intermediate systems with $5\mathrm{d} \leq P_1 < 15$d,
  \item farthest systems with $P_1 \geq 15$d.
\end{enumerate}
For groups 1 and 2, we observe an excess of planets in
external circulation (i.e. with a period ratio higher than the resonant value,
$P_2/P_1 > (p+1)/p$ for the resonance $p+1$:$p$).
However, the excess seems more significant for the closest systems (group 1).
It is also important to notice that there is not any detected close-in system (group 1)
inside the resonance ($P_2/P_1 \approx (p+1)/p$)
whereas a significant number of farther systems (groups 2 and 3) are found in resonance.
Moreover, in the third group, the number of systems inside the resonance is comparable
to or even higher than the number of pairs in external circulation.

\figIII

Figure~\ref{fig:III} shows cumulative distributions of period ratio
in the vicinity of the 2:1 and 3:2 mean-motion resonances for these three groups.
The conclusions are the same as for Fig.\ref{fig:II}.
We performed K-S tests on the distributions given in Fig.~\ref{fig:III}
in order to check the statistical significance of the observed differences between
the three distributions.
The K-S test give the probability to obtain distributions at least as different as the observed ones
with random samplings following the same underlying law.
This probability is of $0.08\%$ for groups 1 and 3.
It is thus very unlikely that both empirical distributions come from the same underlying law and
are this different just by chance.
When comparing the intermediate group (2) with both extreme ones (1 and 3), the differences are
of course less significant and the probabilities given by the K-S test are
respectively $3.5\%$ (groups 1 and 2)
and $10\%$ (groups 2 and 3).

Therefore, we conclude that the distance to the star does have a
statistically significant impact on the distribution of period ratio
of planet pairs.
Very close-in systems ($P_1<5$d) are not found in resonance
and are very often found in external circulation,
whereas for the farthest systems ($P_1\geq 15$d), both populations (resonance and external circulation)
are equivalent with a slight excess of systems inside the resonance.
These observations are well explained by the tidal dissipation scenario of formation of Kepler near-resonant
planets.
On the contrary, the other proposed mechanisms do not predict this dependency on the distance to the star.

\section{Conclusion}

In this letter, we show that the distribution of period ratio among pairs of planets depends
on the distance of the planets to the star.
For close-in systems there is not any detected planet pairs in first-order mean-motion resonances (2:1, 3:2),
and there is an excess of planets in external circulation, i.e. close to the resonance but with a period ratio
higher than the resonant value.
For the farthest systems, the number of resonant pairs is slightly greater than
the number of planets in external circulation.
Using a K-S test to compare both distributions,
we obtain a p-value of $0.08\%$ and conclude that the differences we observe are statistically
significant.
Tidal dissipation raised by the star on the planets naturally explains these observations because
this effect has an important dependency on the distance to the star and is much stronger for close-in systems.
Moreover, it is the only proposed mechanism of formation of these near-resonant systems that predicts such
a strong dependency.

These observations together with the new scenario of formation we proposed
recently \citep[still involving the tidal dissipation but with a faster
evolution of the period ratio, see][section 5]{delisle_resonance_2014}
favor a large influence of
tidal dissipation at the origin of the excess of planets
in external circulation in the \textit{Kepler} data.

\begin{acknowledgements}
  We thank Stéphane Udry for useful advice.
  This work has been supported by PNP-CNRS, CS of Paris
  Observatory, and PICS05998 France-Portugal program.
\end{acknowledgements}

\bibliographystyle{aa}
\bibliography{DL}


\end{document}